\documentclass[10pt,onecolumn]{emulateapj}
\usepackage{graphicx}
\usepackage{bm}
\def\chandra{\it Chandra\rm}
\def\xmm{\it XMM-Newton\rm}
\def\a3112{\it Abell~3112\rm}
\def\rosat{\it ROSAT\rm}
\def\as1101{\it Abell~S1101\rm}
\def\euve{\it EUVE\rm}
\def\sax{\it BeppoSAX\rm}

\def\HI{\it H~I\rm}
\begin{document}

\title{Soft and hard X-ray excess emission in
Abell~3112 observed with Chandra}

\author{Massimiliano Bonamente\altaffilmark{1,2},
Jukka Nevalainen\altaffilmark{3} 
and Richard Lieu\altaffilmark{1}
}
\altaffiltext{1}{Dept.\ of Physics, University of Alabama in Huntsville, Huntsville, AL}
\altaffiltext{2}{Dept.\ of Space Science - NSSTC, NASA Marshall Space Flight Center, Huntsville, AL}
\altaffiltext{3}{University of Helsinki, Helsinki, Finland}

\begin{abstract}
Chandra ACIS-S observations of the galaxy cluster \a3112\  feature
 the presence of an excess of X-ray emission above the contribution from the
diffuse hot gas, which
can be equally well modeled
with an additional non-thermal power-law model or with a low-temperature
thermal model of low metal abundance.
We show that the excess emission cannot be due to
uncertainties in the background subtraction or in the
Galactic \HI\  column density. 
Calibration uncertainties in the ACIS detector that may affect our
results are addressed by comparing the \chandra\ data to \xmm\ MOS and PN
spectra. While differences between the three instruments
remain, all detect the excess in similar amounts, providing
evidence against an instrumental nature of the excess.
Given the presence of non-thermal radio emission
near the center of \a3112, we argue that the excess X-ray emission
is of non-thermal nature and distributed throughout the
entire X-ray bandpass, from soft to hard X-rays.
The excess can be explained with the presence of
a population 
of relativistic electrons with $\sim$7\% of the
cluster's gas  pressure. 
We also discuss a possible thermal nature
of the excess, and examine the problems associated with such
interpretation.

\end{abstract}

\section{Introduction}

The detection of excess extreme-ultraviolet and soft X-ray photons
($\sim$0.1-1 keV) in the spectra of galaxy clusters indicates
the presence of non-thermal processes, or of  warm gas ($T \sim 10^6-10^7$ K) 
near clusters \citep[e.g.,][]{lieu1996,bowyer1996,sarazin1998,
nevalainen2003,bonamente2005}.
The excess emission above the contribution
from the hot intra-cluster medium  was originally discovered 
with \euve\ \citep{lieu1996}, and confirmed by \rosat\ \citep[e.g.,][]{bonamente2002},
 \sax\ \citep[e.g.,][]{bonamente2001b} and \xmm\ \citep[e.g.,][]{nevalainen2003}
for a large number of clusters.
Some detections were subject to criticism ranging from
issues with the \euve\ data analysis \citep{bowyer1999} to
effects of the Galactic \HI\ absorption \citep{bregman2003}
and of the background \citep{bregman2006}; such critiques were
addressed by dedicated re-observations and re-analyses 
which confirmed the presence of the excess 
\citep[e.g.,][]{lieu1999a,bonamente2002,nevalainen2007}.
A tentative detection of emission lines associated with the soft emitter
was reported by \citet{kaastra2003} using \xmm\ data.
Recently, \citet{werner2007} confirmed the soft excess emission
in \as1101\ using {\it Suzaku}
data, but did not confirm the earlier
finding of emission lines associated with the excess emitter.
The presence of hard X-ray excess emission in some 
clusters \citep[e.g.,][]{fusco-femiano1999,nevalainen2004} provides further indication
that additional physics is required in order to interpret 
X-ray cluster spectra.

Interpretation of the soft excess emission has been the subject of
an active debate. 
Thermal emission from warm gas ($kT \sim 0.1-1$ keV)
is a viable model \citep[e.g.,][]{lieu1996,bonamente2002,finoguenov2003},
which indicates the presence of an additional phase in the intergalactic medium
\citep[e.g.][]{cheng2005}.
The thermal interpretation is not consistent with the \citet{cen1999} model of
diffuse filaments, which are too tenuous to 
produce the observed radiation  \citep{mittaz2004,bonamente2005}.
Another viable model is non-thermal emission as Inverse-Compton scattering of the Cosmic
Microwave Background \citep[e.g.][]{sarazin1998,lieu1999b}. 
This interpretation requires the presence of relativistic electrons with a total pressure
generally less than that of the hot gas \citep{lieu1999b,bonamente2005}.

Given that all major X-ray missions with soft X-ray sensitivity since \euve\ 
 have reported a detection of cluster soft excess,
 in this paper we investigate the presence of the
phenomenon in the \chandra\ data of \a3112, a cluster for which \xmm\ detected
the presence of strong soft X-ray excess emission \citep{nevalainen2003}.
The investigation is made possible by the recent calibration efforts
by the \chandra\ team to correct the effects of the contaminant on the optical filter
of ACIS.
The scope of this paper is primarily that of assessing if, with the current calibration
of the soft X-ray channels of ACIS-S, \a3112\ has evidence for the soft excess,
above the calibration uncertainties, as reported by \xmm. 
This paper is structured as follows: in Section \ref{data} we present the
\chandra\ observations of \a3112, in Section \ref{reduction} the
reduction and analysis of the observations with particular attention to removal
of periods of high background, in Section \ref{spectral} we present the
spectral analysis of the \a3112\ data, revealing the presence
of the excess emission, including a comparison
between \chandra\ and \xmm\ spectra. In Section
\ref{background-projection} we discuss the effect of the instrumental background
and of projection effects, and in
Section \ref{interpretation}
we report our interpretation of the spectral analysis.
 Section \ref{conclusions} contains our discussion and conclusions.

\a3112\ is a southern cluster located near R.A.=03h17m52.4s, Dec.=-44d14m35s (J2000) 
and at redshift $z=0.075$,
with an X-ray luminosity of $L_X=7.4 \times 10^{44}$ erg s$^{-1}$ in the 0.1-2.4 keV
band \citep{reiprich2002}. In this paper we assume a cosmology of 
$h=0.72$,
$\Omega_m=0.3$ and $\Omega_{\Lambda}=0.7$, for which 1' corresponds to 
approximately 85 kpc. The Galactic \HI\ column density towards this cluster was
measured by \citet{dickey1990} as $N_H=2.6\times 10^{20}$ cm$^{-2}$.

\section{Chandra observations of Abell~3112}
\label{data}
\chandra\ observed \a3112\ in two separate exposures in September 2001 (observation
ID 2516, 16.9 ks exposure time) and May 2001 (observation ID 2216,
7.2 ks exposure).
The two \chandra\ observations of \a3112\ analyzed in this paper 
 were first published by \citet{takizawa2003}. They found that 
the putative cooling-flow gas in \a3112\ is present
in more modest amounts (44.5 $\pm^{52.1}_{32.5}$ M$_{\odot}$ yr$^{-1}$) 
than previously thought based on \rosat\
data \citep{allen1997,peres1998}. In each of the annuli 
investigated by \citet{takizawa2003} the cooling component is
detected with low significance, and is certainly confined to
the central $\sim 60$" region. \citet{takizawa2003} 
detected the presence of the  central  source PKS~0316-444, 
a radio source detected by the same authors at 1.4 GHz,
which features thermal and non-thermal X-ray emission. 
For these reasons, in this paper
we do not investigate the
diffuse emission in the central 60" region.

Our study of the diffuse X-ray emission from \a3112\  differs
from that of \citet[][who use the 0.5-10 keV band]{takizawa2003} in that more accurate
calibration information is now available, which results in a
better correction for the 
 time-variable
charge transfer inefficiency and  the 
spatially-dependent build-up of contaminants in the
optical filter of the instrument. 
We chose not to use energies below $0.5$ keV, given that
the effects of the optical filter contaminant are still not well calibrated at these
energies (A. Vikhlinin, private communication). 
The soft excess detected in several clusters
becomes stronger at lower X-ray energies; this study therefore probes the presence
of the excess emission only in those channels allowed by the current \chandra\ calibration.

It is worthwhile to point out that \citet{takizawa2003} did detect a sub-Galactic
column density towards \a3112, an effect which may be indicative of excess
soft X-ray photons, as shown in \citet{bonamente2005} 
and \citet{nevalainen2007}. They attributed this effect to an over-correction
of the \chandra\ effective area by the \chandra\ data analysis tools available at the time of 
their study. 
 
\section{Data reduction}
\label{reduction}
The data reduction was performed with CIAO  3.4, using the
calibration information available in CALDB  3.3.~\footnote{The analysis
was initially performed using CIAO  3.3 and
of CALDB 3.2. The later release of the \chandra\ CALDB
included changes to the calibration of the charge transfer inefficiency 
(CTI) for back-illuminated chips; such changes 
did affect the \a3112\ spectra, but 
did not change the  overall results presented in this paper.}
Level 1 event files were reprocessed to apply the latest calibration
(using the \verb acis_process_events \ tool), including
the time- and space-dependent correction due to the contaminant on the ACIS
optical filter, which affects the detection of soft X-ray photons.
ACIS observations of bright sources can also be affected by a readout artifact
also known as out-of-time events \citep[e.g.,][]{markevitchscript}. It is caused
by source photons that hit the detector during the $\sim 40$ ms that
are necessary for one ACIS frame (accumulated over a $\sim 3.2$ s integration) to be
transferred to the readout electronics. Although our observations feature
neither a strong point source nor a peaked distribution of the cluster surface
brightness, we follow the additional reduction step described by \citet{markevitchscript} 
in order to account for this effect.


The ACIS-S3 background was studied in detail by \citet{markevitch2003}, who
provide a detailed set of prescriptions useful to excise times with high background
count rates.
Since the cluster occupies the entire
S3 chip, the S1 chip was used to investigate the presence of background flares.
Following the \citet{markevitch2003} prescription, 
the quiescent background rate in the 2.5-6.5 keV band was determined from the 
flare-free blank-sky dataset 
provided with CIAO, and applicable to the A3112 observations 
(\verb acis57sD2000-12-01bkgrndN0002.fits ). 
Light curves in 1 ks bins  were extracted from source-free regions of the S1 chip, and
time intervals with count rates above and below 20\% of the blank-sky mean 
($1.46 \times 10^{-3}$ counts s$^{-1}$ arcmin$^{-2}$) were  discarded.
This procedure resulted in 8.7 ks (Observation 2516) and 4.5 ks (Observation 
2216) of clean data.

Both \a3112\ observations and the  blank-sky dataset
were obtained in \verb VFAINT \ mode, an ACIS detector mode which enables a
reduction of events likely associated with cosmic rays.
Spectra and response files were generated using the
\verb specextract \ CIAO tool. 

A spectrum was extracted for the region 1-2.5' of each observation
around the centroid of the X-ray emission.
Our choice  was determined
by the needs of (a) excising the central region where cooler
gas may be present, (b) avoiding outer regions where the background
becomes dominant and (c) accumulating a spectrum with sufficient
number of counts in order to perform the spectral analysis; 
the short observing time did not
allow a study with finer spatial resolution. 
After ensuring that the two observations were consistent with each other,
the two spectra were added, and the response files properly averaged.
The spectrum was also rebinned in order 
to ensure that at least 25 counts are present in 
each bin.

\section{Spectral analysis }
\label{spectral}
The ACIS spectrum is initially fit to an optically-thin plasma emission code
(\verb mekal \ in XSPEC), modified by 
the photoelectric absorption (\verb wabs \ in XSPEC); the Galactic $N_H$ 
was fixed
at the measured values,  except in a model in Section \ref{singletemp} in which
a variable $N_H$ is explicitly stated~\footnote{
The normalization of the spectra
in Table \ref{table-1mekal}, \ref{table-1mekal_other}, \ref{table-mekal-po} and \ref{table-mekal-mekal}
follows the customary XSPEC units, $N=\frac{10^{-14}}{(1+z)^2 D_A^2} \int n_e n_H dV$, where $D_A$ is
the angular diameter distance (cm), and $n_e$, $n_H$ are the electron and hydrogen densities
(cm$^{-3}$) respectively.}. The background is measured from blank-sky observations,
as described in Section \ref{reduction}.
The spectral range used in this paper for the ACIS data is 0.5-7 keV.
Errors are 68\% confidence intervals, obtained with the $\chi^2_{\rm min}+1$ method.

In order to assess the possible impact of calibration uncertainties on the
\chandra\ data analysis, we also analyze \xmm\ MOS and PN spectra
of the same region of \a3112. The \xmm\ data analysis is described in detail in \citet{nevalainen2003},
to which the reader is referred for details. We reduced the available 
\xmm\ observation of \a3112\ to obtain 22.3 ks of MOS data, and 16.6ks of
PN data. The \xmm\ data analysis was performed with the SAS 7.0.0 software, using
the calibration information available as of May 2007. The \xmm\ reduction follows
the same steps as the \chandra\ data, including flagging of high-background
time intervals, and the use of background accumulated from blank-sky observations
\citep[see][]{nevalainen2005}.
Data from the two MOS units were averaged to yield one MOS spectrum.
The \xmm\ spectra are fit in the 0.3-8 keV band.

\subsection{Single temperature model }
\label{singletemp}
\begin{figure}
\includegraphics[width=2.5in,angle=-90]{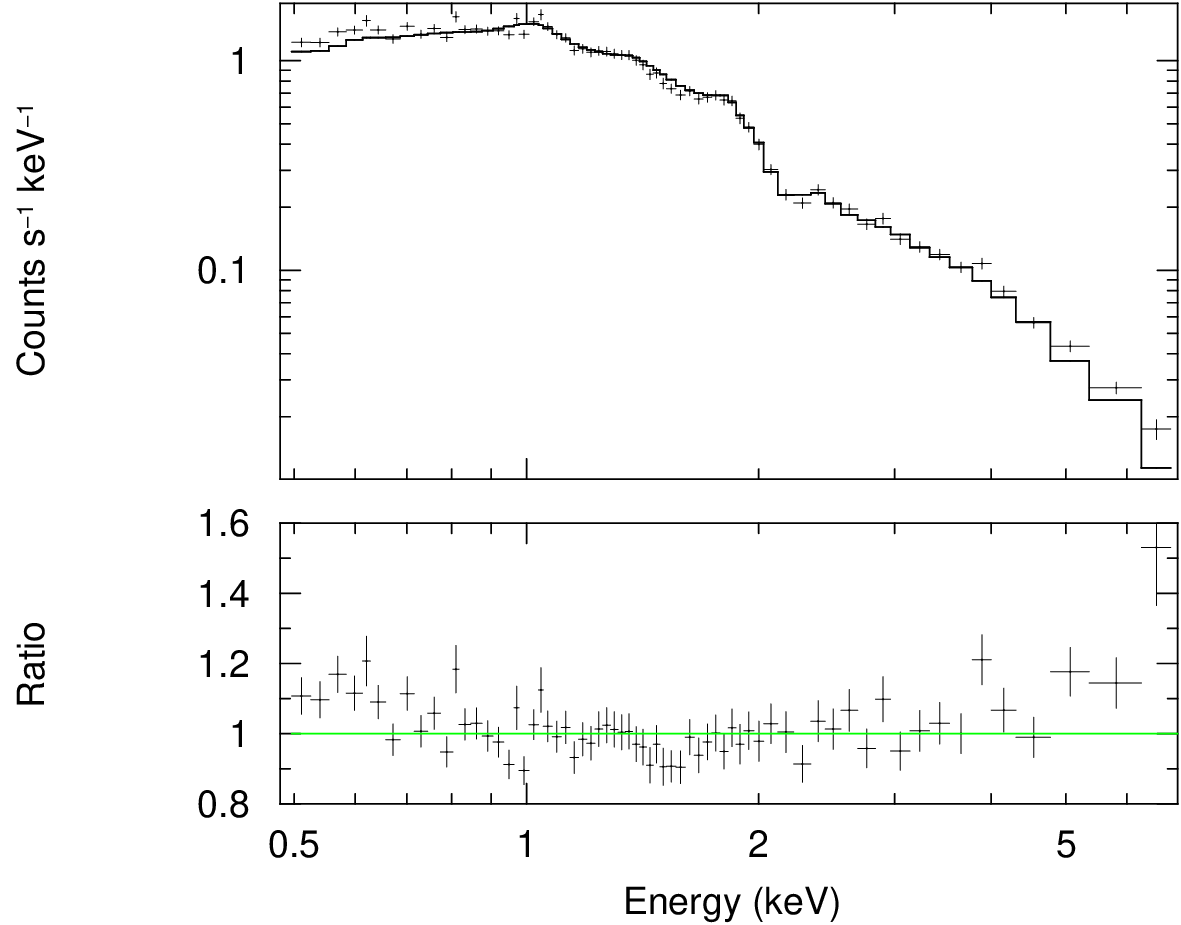}
\includegraphics[width=2.5in,angle=-90]{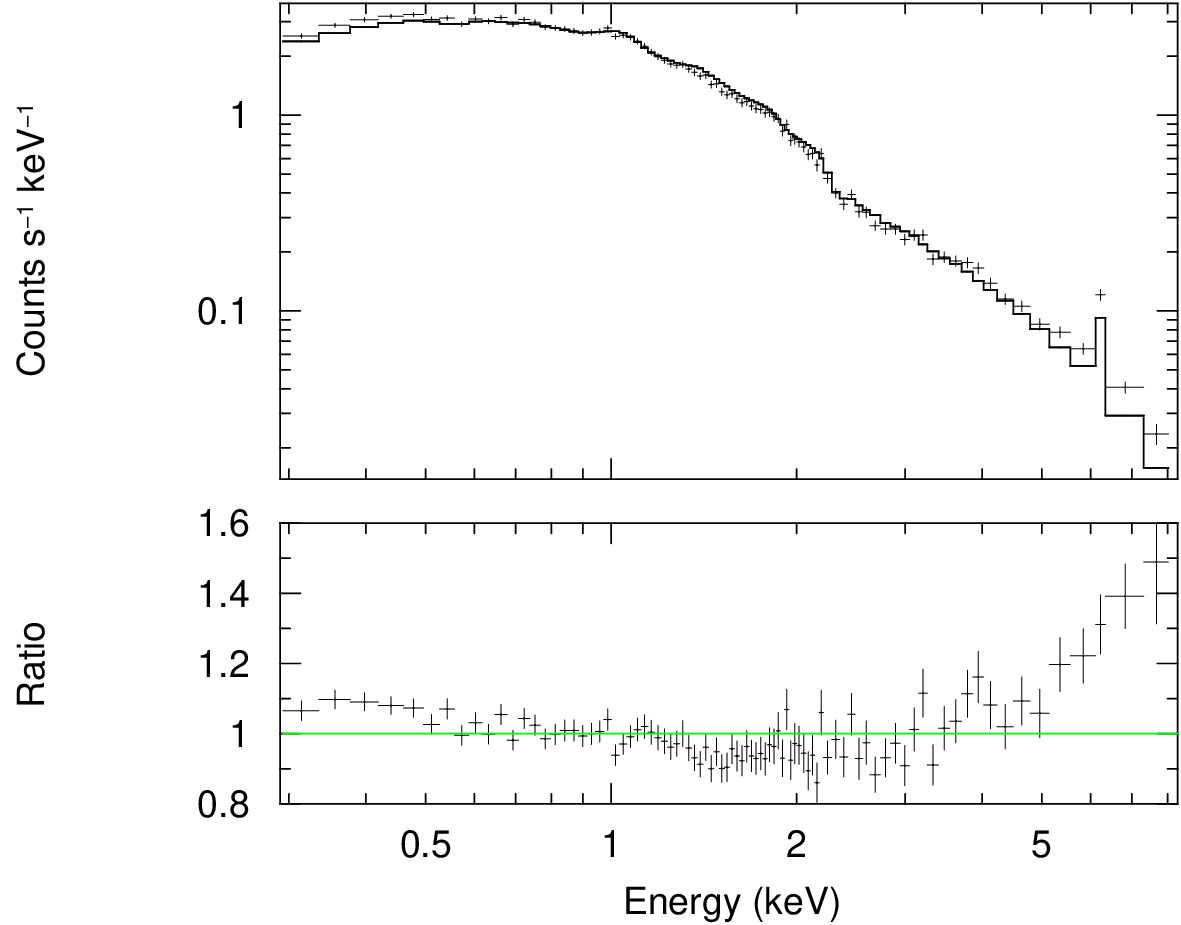}

\includegraphics[width=2.5in,angle=-90]{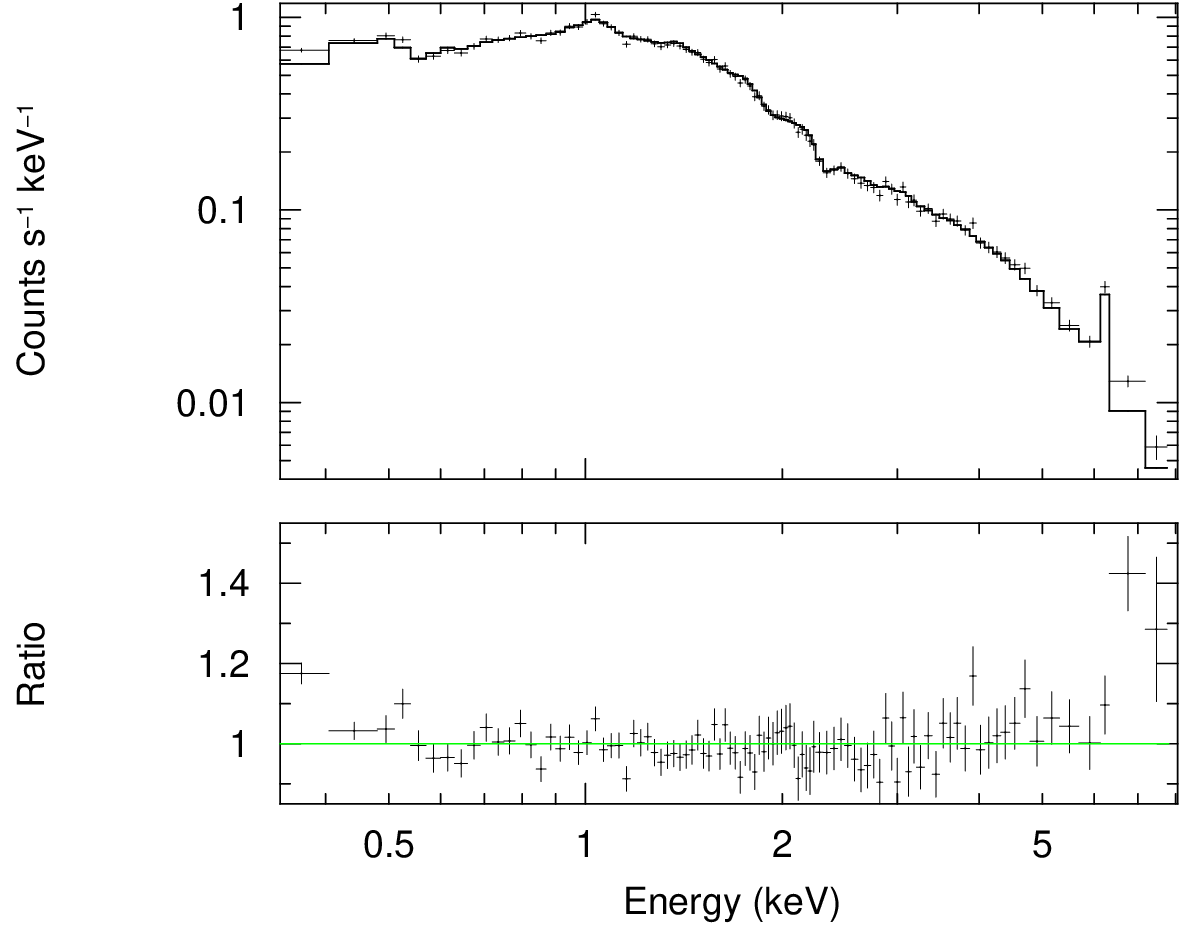} 
\includegraphics[width=2.5in,angle=-90]{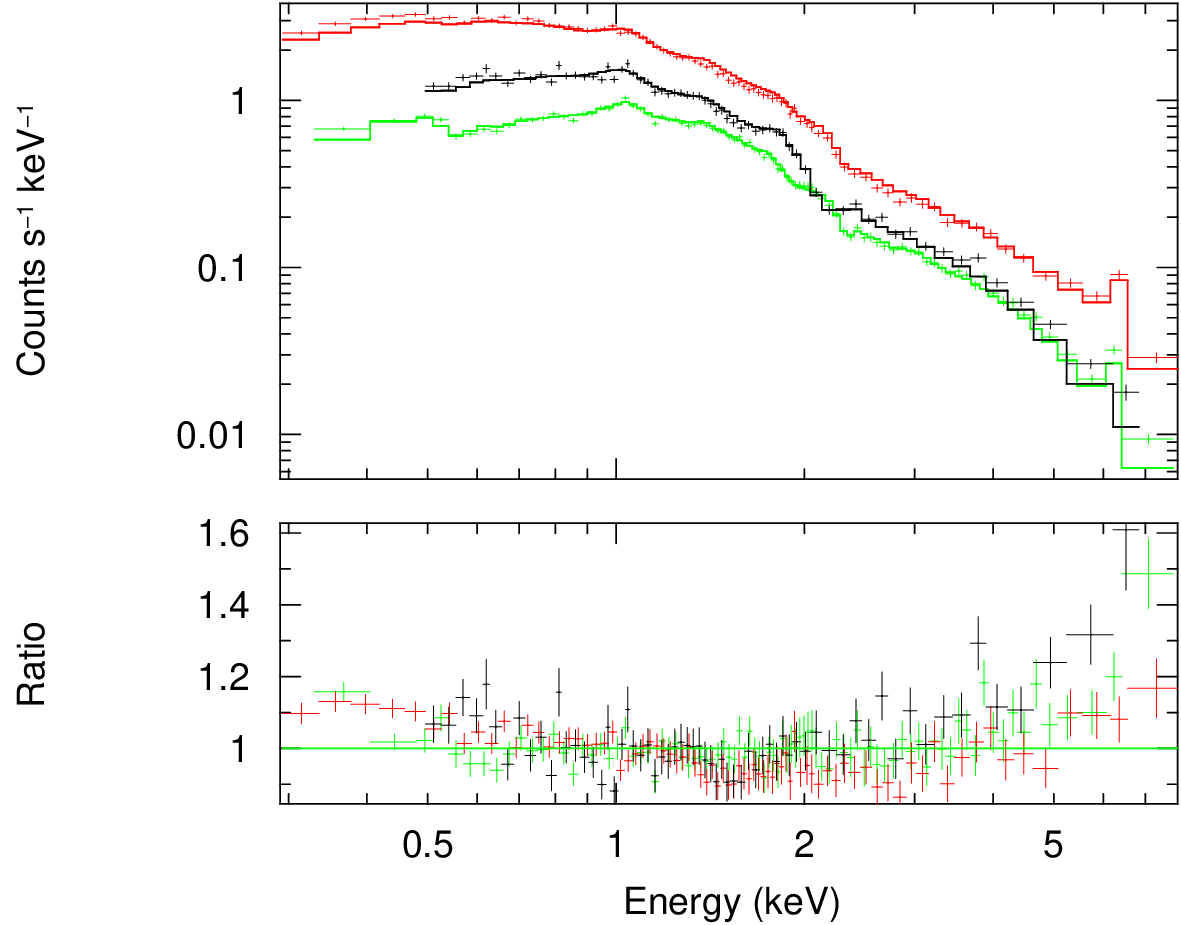}
\caption{Fit of \a3112\ data to a single temperature model. 
Top, from left to right: 
 ACIS and PN; Bottom, from left to right: MOS and joint ACIS/PN/MOS fit (black: ACIS;
red: PN; green: MOS).
The ACIS data were fit to the 0.5-7 keV band, the PN and MOS data were
fit to the 0.3-8 keV band.
\label{1mekal}}
\end{figure}

\begin{figure}
\includegraphics[width=2.5in,angle=-90]{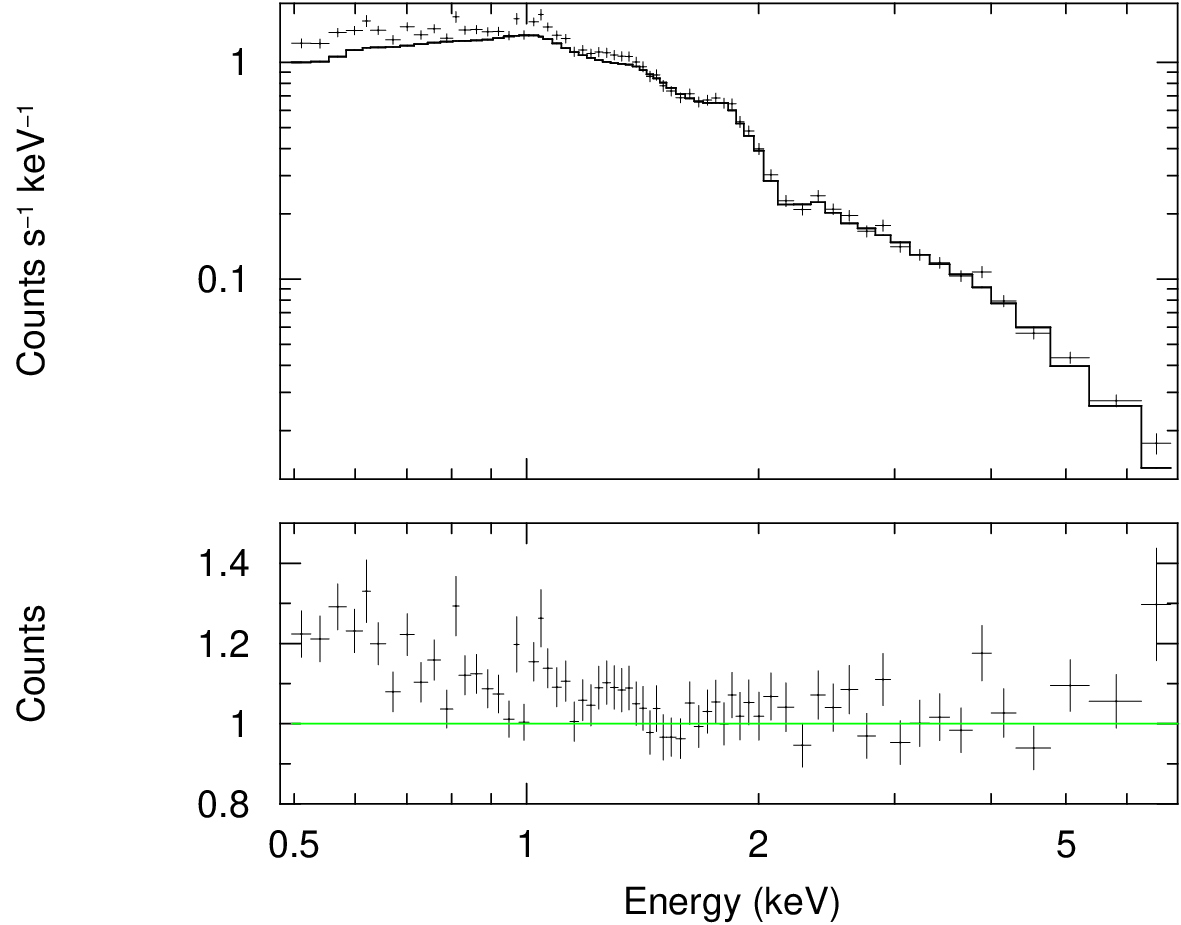}
\includegraphics[width=2.5in,angle=-90]{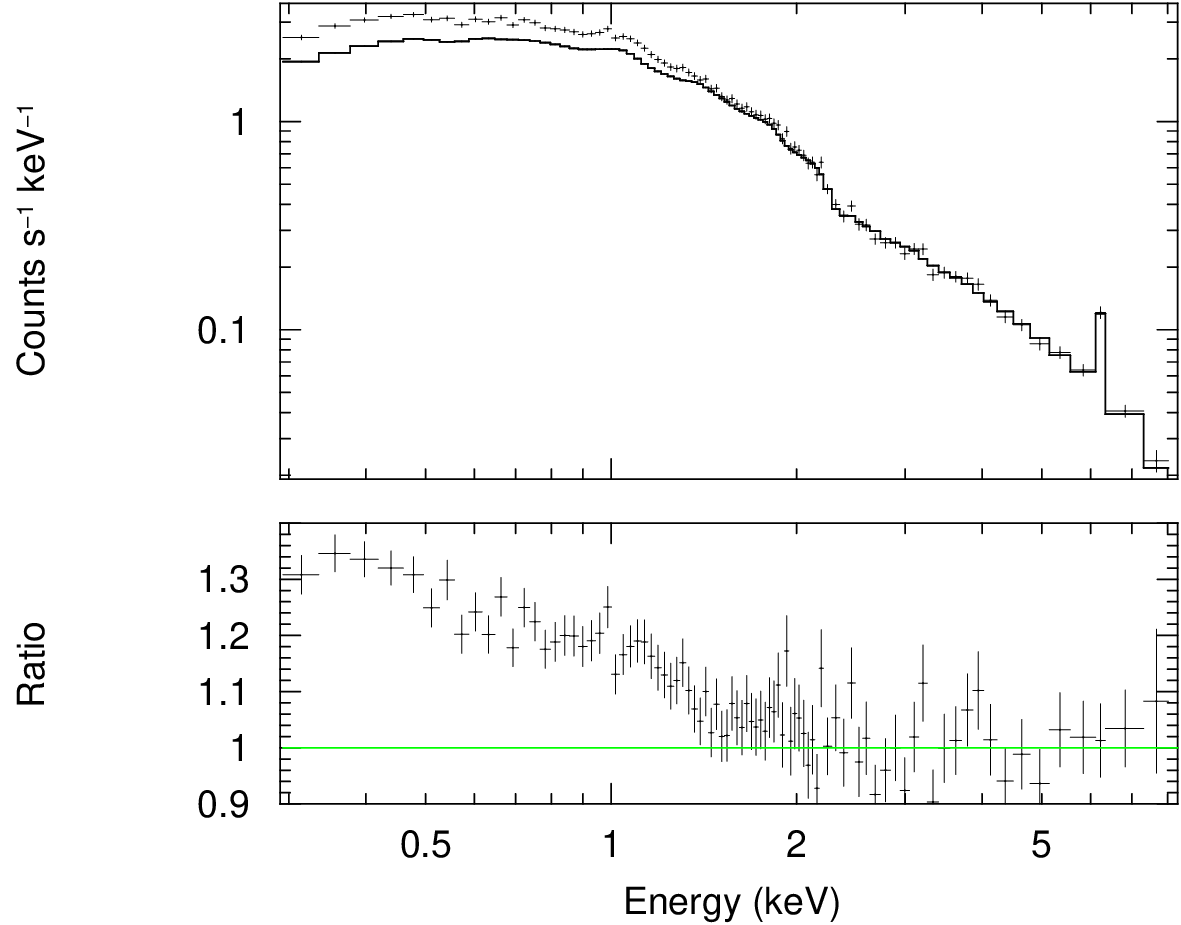}

\includegraphics[width=2.5in,angle=-90]{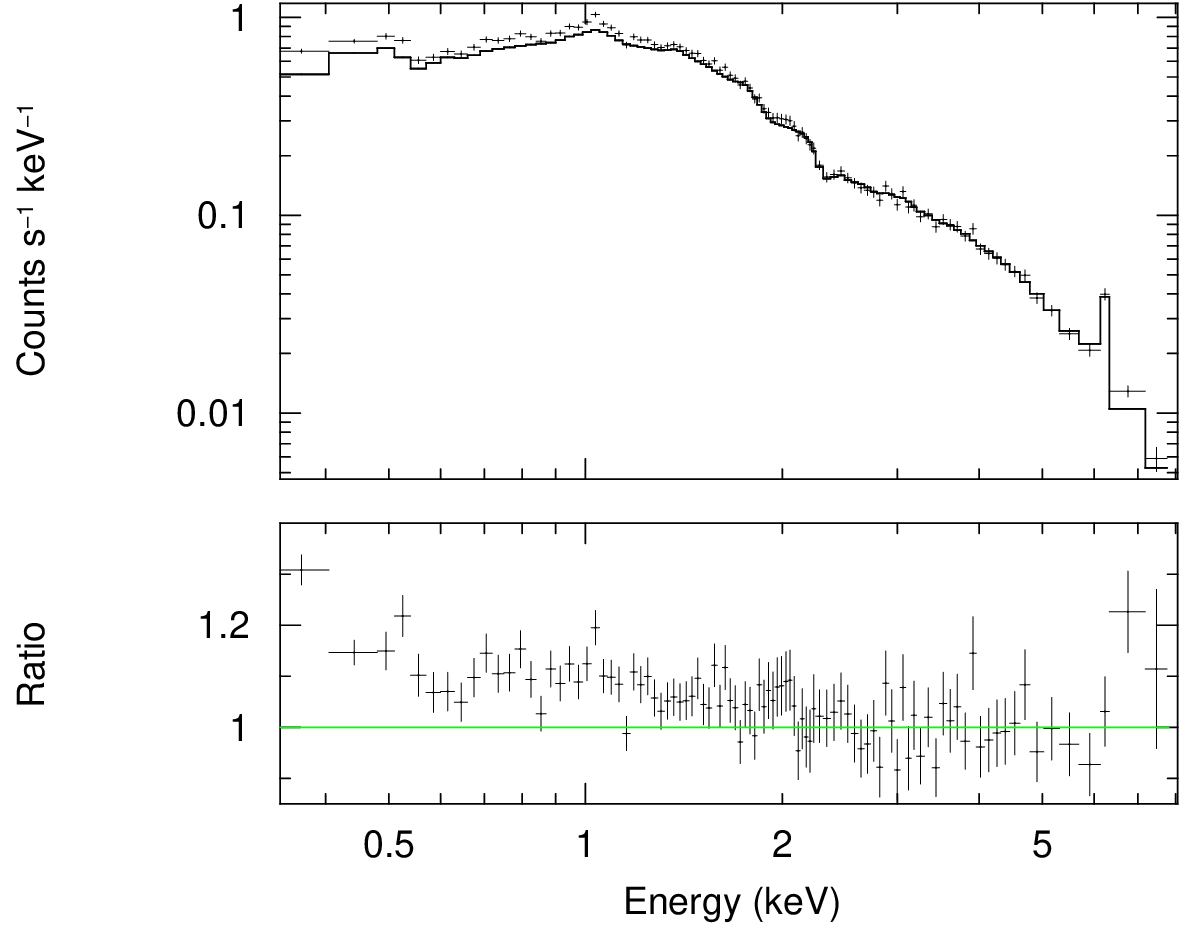}
\includegraphics[width=2.5in,angle=-90]{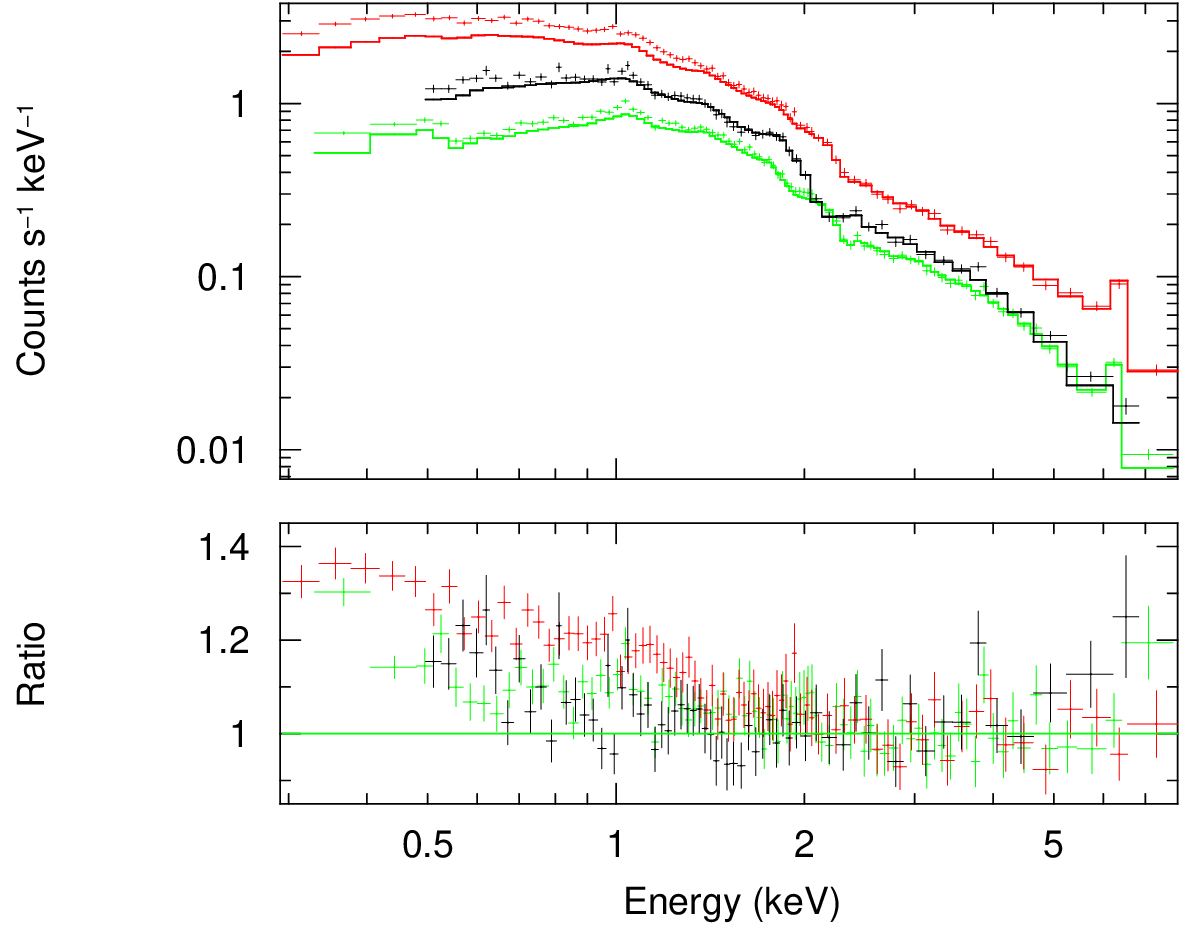}

\caption{Fit of \a3112\ data to a single mekal model in 
the hard band, 
and extrapolation of the model to low energy.
Top, from left to right:
 ACIS and PN; Bottom, from left to right: MOS and joint ACIS/PN/MOS fit  (black: ACIS;
red: PN; green: MOS).
The ACIS data were fit to the 2-7 keV band, the PN and MOS data were 
fit to the 2-8 keV band.
\label{1mekal_2-7}}
\end{figure}

\begin{figure}
\includegraphics[width=2.5in,angle=-90]{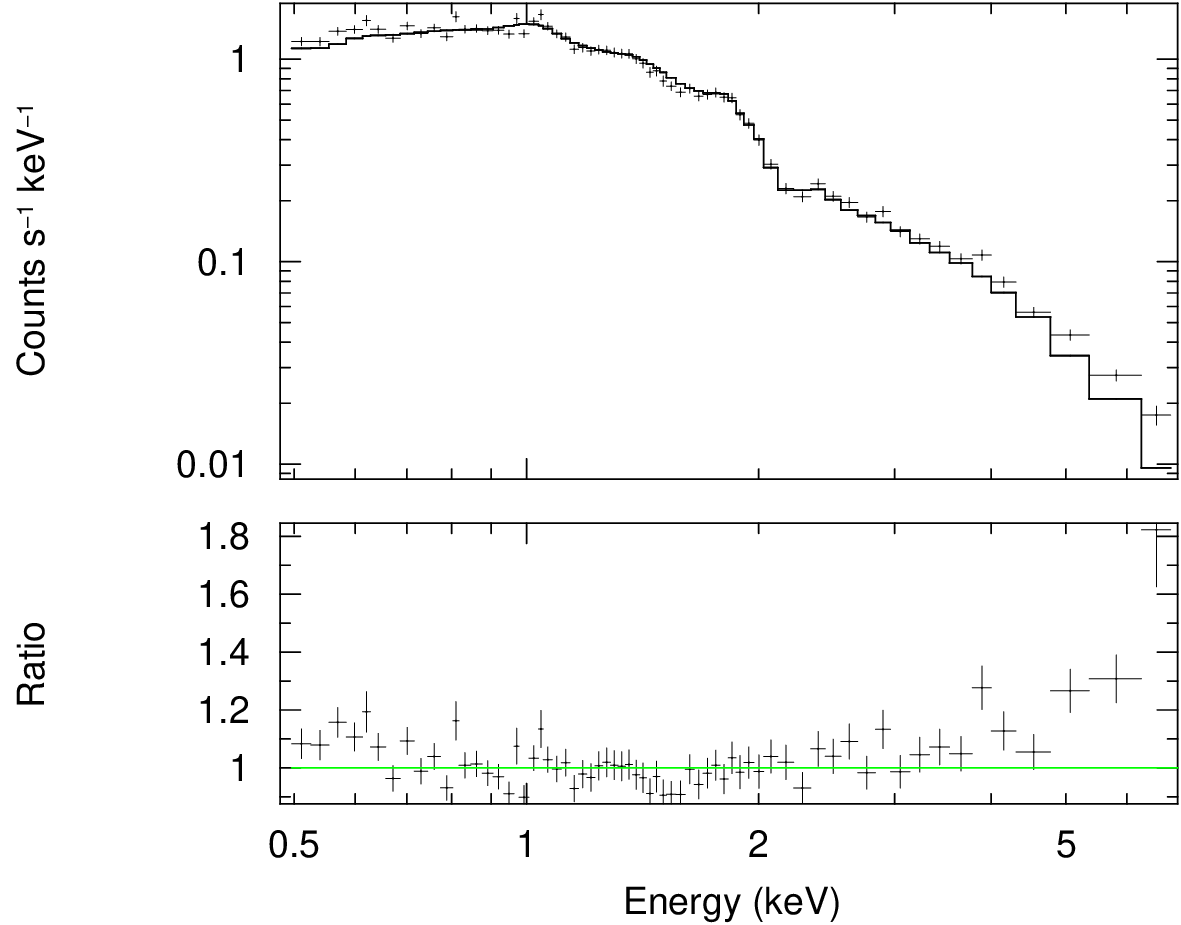}
\includegraphics[width=2.5in,angle=-90]{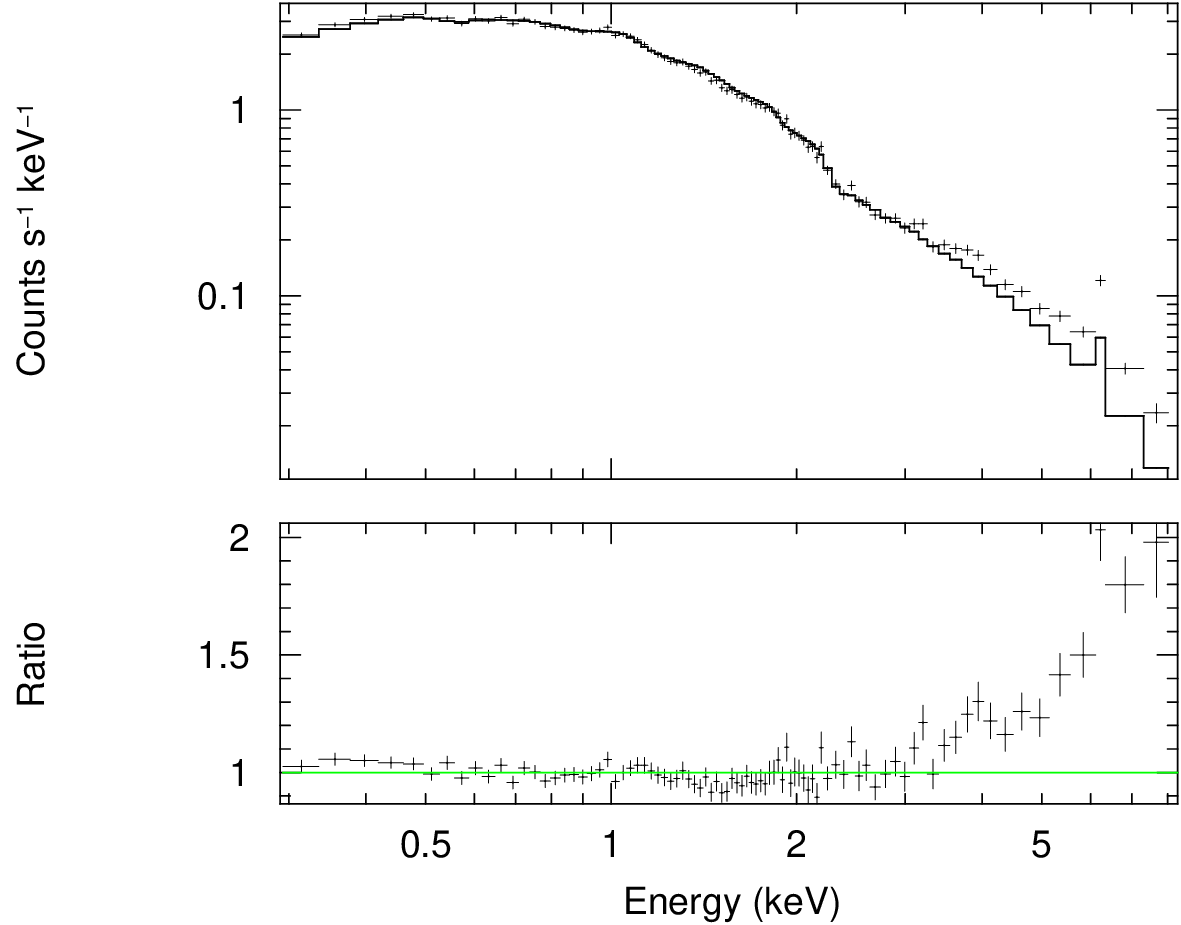}

\includegraphics[width=2.5in,angle=-90]{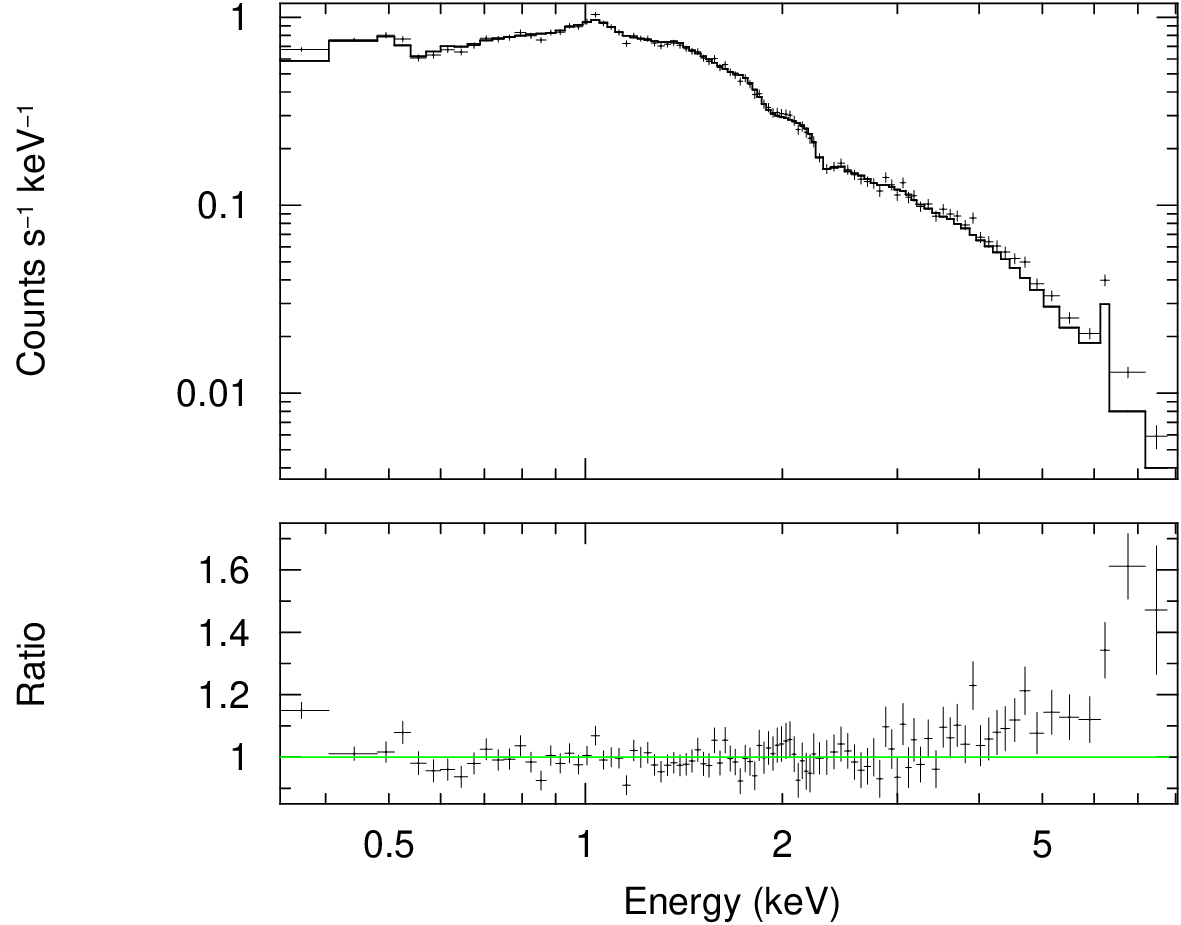}
\includegraphics[width=2.5in,angle=-90]{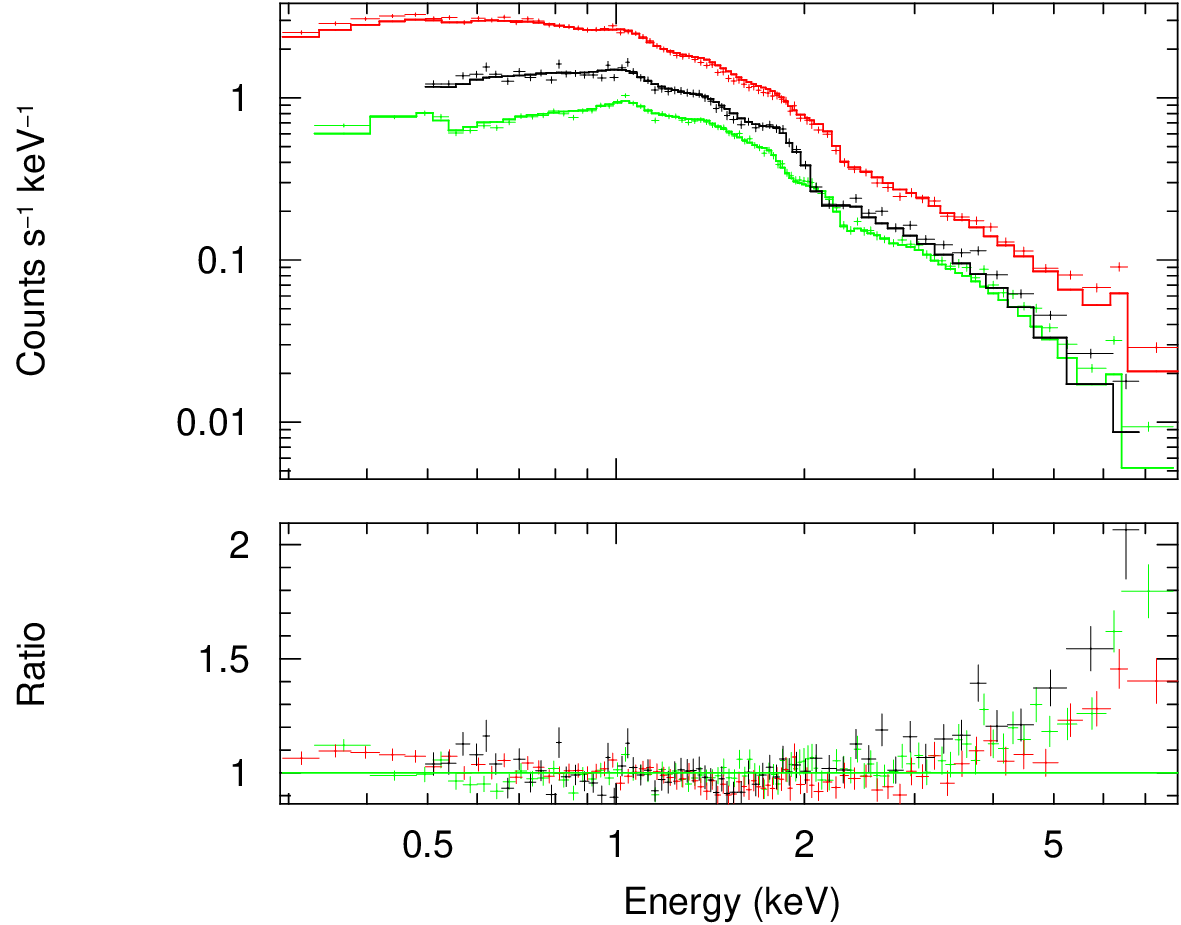}
\caption{Fit of \a3112\ data to a single mekal model in the soft band,
and extrapolation of the model to the hard band.
Top, from left to right:
 ACIS and PN; Bottom, from left to right: MOS and joint ACIS/PN/MOS fit (black: ACIS;
red: PN; green: MOS).
The ACIS data were fit to the 0.5-4 keV band, the PN and MOS data 
were fit to the 0.3-4 keV band.
\label{1mekal_0.5-4}}
\end{figure}

First, we fit the ACIS spectrum with a simple one-temperature model in the 0.5-7 keV band.
The fit  is poor, with positive residuals at
low and high energy, and negative residuals in the central band around 2 keV
(Fig. \ref{1mekal} and Table \ref{table-1mekal}).
The presence of residuals is such that a fit of the spectrum in the hard band
(2-7 keV) provides a significantly higher temperature than the 0.5-7 keV fit,
and its extrapolation to low energies does not match the observed spectrum
(Fig. \ref{1mekal_2-7} and Table \ref{table-1mekal_other}).
Likewise, a fit to the low-energy band alone (0.5-4 keV) provides a lower
temperature, and highlights the presence of high-energy residuals
(Fig. \ref{1mekal_0.5-4} and Table \ref{table-1mekal_other}). 
We provide similar fits to the \xmm\ spectra, in which the 0.3-8 keV band is used for the
full-band fit, and the 2-8~keV and 0.3-4~keV bands respectively for the low-energy and
high-energy band fits. Finally, a joint fit to all available data
(ACIS, MOS and PN) is also performed for comparison.~\footnote{The 
effective areas of the \xmm\ instruments
differ from that of ACIS; in particular, 
ACIS's effective area curve falls off more steeply
at low and high energies than those of MOS and PN,
and PN has an average effective
area that exceeds both those of ACIS and MOS by a factor of few. These differences 
in instrumental response
result in different sensitivity to X-ray photons for the three instruments,
and therefore even a comparison of these instruments over the same formal band (say, 0.5-7 keV) 
would not be exact. The wider \xmm\ bandpass we use enables a better identification
of low and high energy residuals, and the spectra in Figures \ref{1mekal}-\ref{1mekal_0.5-4}
enable a detailed comparison in each spectral bin between the two observatories.}

In Section \ref{projection} we show that the poor fit to a single temperature model
cannot be due to the decrease of the hot gas temperature at large radii.
From these narrow-band fits alone it is not possible to establish whether the 
poor single-temperature fit is due to an additional high-energy 
component (e.g., a hard excess) or to a low-energy component (e.g., a soft excess).
What these fits do indicate is the need for a more accurate modeling
than  a simple single temperature model, if one wants to satisfactorily fit
the whole-band spectrum.

\begin{deluxetable}{lccccccccc}
\tabletypesize{\scriptsize}
\tablecaption{Single temperature model with fixed and variable $N_H$\label{table-1mekal}}
\startdata
\hline
 & \multicolumn{4}{c}{Fixed Galactic $N_H=2.6\times 10^{20}$ (cm$^{-2}$)} &  \multicolumn{5}{c}{Free $N_H$} \\
 & \multicolumn{4}{c}{\hrulefill} & \multicolumn{5}{c}{\hrulefill} \\
Data & $\chi^2$/dof ($\chi^2_r$) & $kT$ & $A$ & Norm & $\chi^2$/dof ($\chi^2_r$) & $kT$ & $A$ & Norm & $N_H$\\
                &    &  (keV) & & ($\times 10^{-2}$) & & (keV) & & ($\times 10^{-2}$) & $(10^{20}$ cm$^{-2}$)\\
\hline
ACIS & 345.7/269 (1.29) & 4.94$\pm$0.14 & 0.40$\pm$0.05 & 1.01$\pm$0.01 & 
       294.9/268 (1.10) & 5.70$\pm^{0.15}_{0.14}$ & 0.50$\pm$0.06 & 0.94$\pm$0.01 & $\leq$0.35 \\
\multicolumn{9}{c}{\dotfill}\\
PN   & 274.3/161 (1.70) & 3.97$\pm^{0.10}_{0.05}$ & 0.23$\pm$0.02 & 0.97$\pm$0.01 & 
       155.3/160 (0.97) & 4.47$\pm^{0.06}_{0.07}$ & 0.32$\pm$0.02 & 0.91$\pm$0.01 & 1.10$\pm^{0.2}_{0.1}$ \\
MOS  & 236.6/173 (1.77) & 4.52$\pm^{0.05}_{0.07}$ & 0.34$\pm$0.03 & 1.05$\pm$0.01 &
       189.6/172 (1.10) & 4.94$\pm^{0.07}_{0.14}$ & 0.39$\pm^{0.04}_{0.02}$ & 1.00$\pm$0.01 & 1.34$\pm$0.23\\
Joint$^{\tablenotemark{a}}$& 918.3/606 (1.52) & 4.37$\pm^{0.02}_{0.07}$ & 0.29$\pm^{0.02}_{0.01}$ & 1.04$\pm$0.01 &
       685.7/605 (1.13) & 4.83$\pm^{0.05}_{0.06}$ & 0.38$\pm^{0.02}_{0.03}$ & 0.97$\pm$0.01 & 0.98$\pm$0.13
\\
     &                  &                         &                         & 0.95$\pm$0.01 &
                        &                         &                         & 0.99$\pm$0.01 &  
\\
     &                  &                         &                         & 1.05$\pm$0.01 &
                        &                         &                         & 0.87$\pm$0.01 &
\\

\enddata
\tablenotetext{a}{For the joint fit, the three normalizations apply respectively to the ACIS, PN and MOS data.}
\end{deluxetable}

\begin{deluxetable}{lcccc}
\tablecaption{Narrow-band single temperature models with fixed $N_H$\label{table-1mekal_other}}
\startdata
\hline
Data & $\chi^2$/dof ($\chi^2_r$) & $kT$ & $A$ & Norm  \\
                &     &  (keV) & & $(10^{-2})$  \\
\hline
\multicolumn{5}{c}{\it Fit to 2-7 keV band (2-8 keV for \xmm\ data)}\\
ACIS & 179.7/166 (1.08) & 5.88$\pm^{0.57}_{0.36}$ & 0.45$\pm{0.07}$ & 0.94$\pm{0.03}$\\
\multicolumn{5}{c}{\dotfill}\\
PN & 73.4/105 (0.70) & 5.28$\pm^{0.23}_{0.21}$ & 0.32$\pm^{0.04}_{0.03}$ & 0.82$\pm$0.02 \\
MOS & 115.4/105 (0.95) & 5.34$\pm^{0.22}_{0.19}$ & 0.37$\pm$0.04 & 0.97$\pm$0.02\\
Joint$^{\tablenotemark{a}}$ & 365.6/394 (0.93) & 5.31$\pm^{0.14}_{0.13}$ & 0.36$\pm^{0.03}_{0.02}$ & 0.98$\pm^{0.02}_{0.01}$\\
      &                  &                         &                         & 0.81$\pm$0.01 \\
      &                  &                         &                         & 0.97$\pm$0.01 \\
\hline
\multicolumn{5}{c}{\it Fit to 0.5-4 keV band (0.3-4 keV for \xmm\ data)}\\
ACIS & 261.3/210 (1.24) & 4.54$\pm$0.014 & 0.29$\pm$0.06 & 1.04$\pm$0.02 \\
\multicolumn{5}{c}{\dotfill}\\
PN & 137.0/108 (1.29) & 3.46$\pm^{0.07}_{0.06}$ & 0.13$\pm$0.02 & 1.01$\pm$0.01\\
MOS & 165.6/116 (1.43) & 4.24$\pm$0.08 & 0.26$\pm$0.03 & 1.08$\pm$0.01\\
Joint$^{\tablenotemark{a}}$ & 638.6/437 (1.46) & 3.93$\pm^{0.06}_{0.05}$ & 0.20$\pm^{0.01}_{0.03}$ & 1.07$\pm$0.01\\
      &                  &                         &                         & 0.97$\pm$0.01\\
      &                  &                         &                         & 1.10$\pm$0.01\\
\enddata

\tablenotetext{a}{For the joint fit, the three normalizations apply respectively to the ACIS, PN and MOS data.}
\end{deluxetable}

Before proceeding with multi-component models, 
we investigate the possibility that the poor single-temperature fit 
 is caused by variations in the \HI\ absorbing column, and 
therefore repeat the single-temperature fits with a variable $N_H$.
The fits to a free-$N_H$ single-temperature model are
somewhat improved (Table \ref{table-1mekal}).
The fact that the best-fit
$N_H$ is significantly sub-Galactic, and that 
high-energy residuals remain, indicates that
the fit residuals cannot be explained as a Galactic absorption effect.
We therefore proceed with
the addition of a second emission component in order to provide a better
fit to the data.

\subsection{Two component models: non-thermal model and two-temperature model}

\begin{figure}
\begin{center}
\includegraphics[width=2.2in,angle=-90]{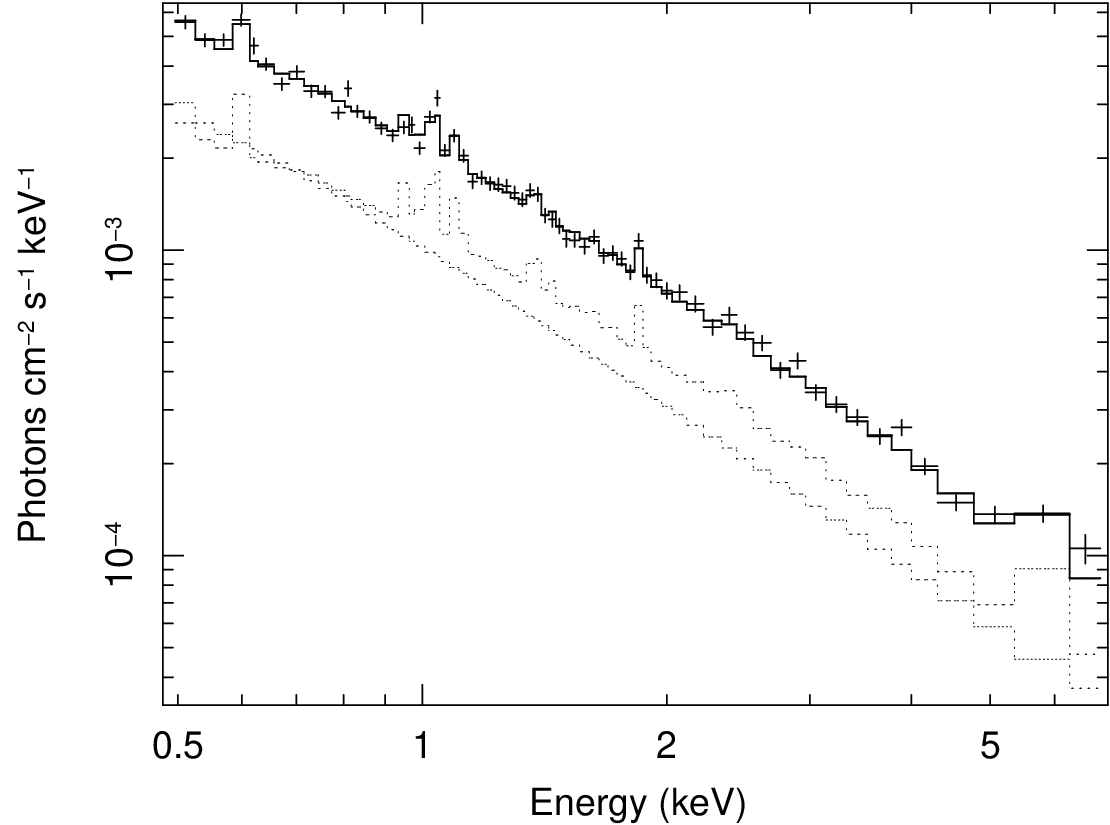}
\includegraphics[width=2.2in,angle=-90]{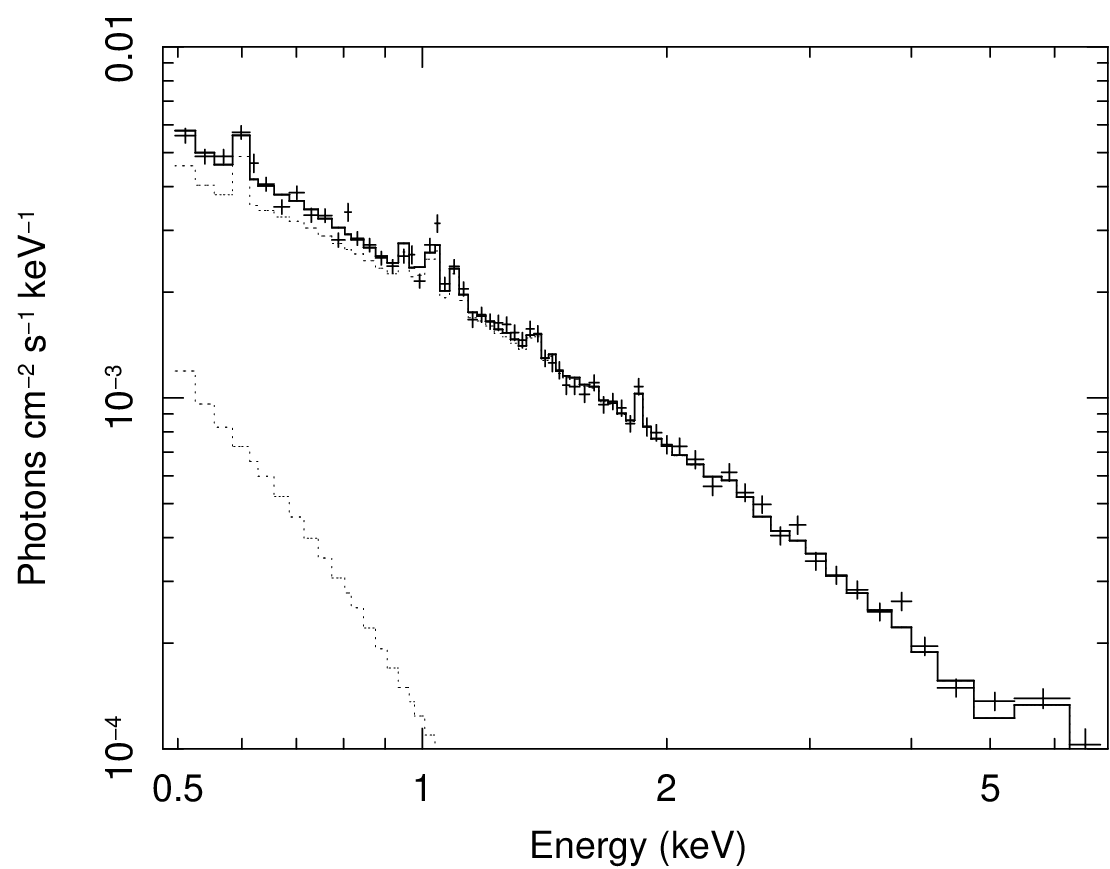}
\end{center}
\caption{Left: Non-thermal fit to ACIS data, dotted
lines are individual model components. Right: Two-temperature fit
to ACIS data.
Models have the parameters of Table \ref{table-mekal-po}}
\label{po}
\end{figure}

We now add a non-thermal power law component to the thermal model. 
The addition results in acceptable fits, with
a significant improvement to the $\chi^2$ statistic 
(Table \ref{table-mekal-po}).
The power-law model of the 1-2.5' region
contributes
to $\sim$50\% of the X-ray luminosity in the same region
(Fig. \ref{po}).
This results in 
higher metal abundances for the hot gas, now with an 
emission integral reduced by $\sim$50\% with respect to
the single-temperature model. 
If the excess emission is described by this
non-thermal power-law model, the excess emission would extend throughout
the X-ray band and into the hard X-ray band.
\begin{deluxetable}{lcccc|cc}
\tablecaption{Non-thermal  model \label{table-mekal-po}}
\startdata
\hline
Data & $\chi^2$/dof  ($\chi^2_r$) & $kT$ & $A$ & Norm. & $\alpha$ &  $L_{42}$~\tablenotemark{(a)}  \\
       &                   & (keV) &  & ($\times 10^{-2}$) & & ($10^{42}$ erg s$^{-1}$) \\
\hline
ACIS & 195.4/267 (1.10) & 5.36$\pm^{0.43}_{0.56}$ & 0.87$\pm^{0.25}_{0.10}$ & 0.50$\pm^{0.14}_{0.11}$ &
       1.79$\pm^{0.11}_{0.06}$& 74.4 \\
\multicolumn{7}{c}{\dotfill}\\
PN & 118.6/159 (0.74) & 4.12$\pm^{0.24}_{0.23}$ & 0.47$\pm^{0.05}_{0.04}$ & 0.57$\pm^{0.03}_{0.05}$ &
       1.83$\pm^{0.07}_{0.04}$ & 61.0 \\
MOS & 185.1/172 (1.08) & 4.44$\pm^{0.19}_{0.25}$ & 0.48$\pm$0.05 & 0.74$\pm$0.06 & 
       1.74$\pm^{0.07}_{0.06}$ & 49.8 \\
Joint$^{\tablenotemark{b}}$ & 643.7/604 (1.07) & 4.59$\pm^{0.09}_{0.08}$ & 0.50$\pm^{0.03}_{0.02}$ & 0.69$\pm$0.10 &
       1.85$\pm^{0.02}_{0.04}$ & 46.0   \\
      &                  &                         &                         & 0.59$\pm^{0.03}_{0.01}$ &
                               & 56.0 \\
      &                  &                         &                         & 0.71$\pm$0.01 &
                               & 56.0 \\
\enddata

\tablenotetext{a}{$L_{42}$ is the unabsorbed luminosity of the non-thermal model in the 0.5-7 keV band.}
\tablenotetext{b}{For the joint fit, the three normalizations apply respectively to the ACIS, PN and MOS data.}

\end{deluxetable}
Alternatively, we add another thermal component to the model.
The addition of a second thermal component indicates that the
excess emission can be modeled as a diffuse gas 
at lower temperature ($\sim 0.3$ keV) and 
with low metal abundance,
with spectral fits  of similar quality to the non-thermal model (Table \ref{table-mekal-mekal}).
\begin{deluxetable}{lcccc|ccc}
\tablecaption{Two-temperature model \label{table-mekal-mekal}}
\startdata
\hline
Data & $\chi^2$/dof ($\chi^2_r$) & $kT_{\rm hot}$ & $A$ & Norm. & $kT_{\rm warm}$ & $A$  & Norm.\\
   &                       & (keV)   &    & ($\times 10^{-2}$) & (keV)& &($\times 10^{-3}$)\\
\hline
ACIS & 296.6/266 (1.11) & 5.61$\pm^{0.27}_{0.33}$ & 0.51$\pm^{0.07}_{0.05}$ & 0.94$\pm$0.03 &
       0.34$\pm^{0.32}_{0.12}$ & $\leq$2.5 & 0.14$\pm^{0.17}_{0.03}$ \\
\multicolumn{8}{c}{\dotfill}\\
PN & 117.8/158 (0.74) & 5.34$\pm^{0.23}_{0.30}$ & 0.38$\pm^{0.03}_{0.04}$ & 0.78$\pm$0.04 &
       0.89$\pm^{0.20}_{0.16}$ & $\leq$0.17 & 0.9$\pm^{0.04}_{0.03}$ \\
MOS & 178.8/170 (1.05) & 4.68$\pm^{0.10}_{0.07}$ & 0.39$\pm{0.03}$ & 1.03$\pm$0.01 &
       $\leq$0.10 & $\leq$0.04 & 3.39$\pm^{1.59}_{1.02}$ \\
Joint & 654.5/603 (1.09) & 5.12$\pm^{0.09}_{0.05}$ &  0.39$\pm^{0.03}_{0.02}$ & 0.94$\pm$0.01 &
       0.62$\pm^{0.04}_{0.10}$ & $\leq$0.01 &  0.24$\pm^{0.02}_{0.01}$ \\
      &                  &                         &                          & 0.59$\pm$0.01 &
                               &            &                           \\
      &                  &                         &                          & 0.71$\pm$0.01 &
                               &            &                           \\

\enddata
\end{deluxetable}
\subsection{Comparison to XMM-Newton results and effect of
calibration uncertainties}
The \xmm\ study of \citet{nevalainen2003} found soft excess emission
in \a3112\ at the level of $\sim$20-40\% above the hot thermal component.
The analysis of \chandra\ and \xmm\ calibration is an ongoing
effort, and changes to the \xmm\ calibration since the publication
of the \citet{nevalainen2003} resulted in
changes to the derived soft excess properties \citep[also noted by][]{nevalainen2007}, although the
presence of the excess itself was confirmed.

In  order to provide a more up-to-date comparison between the \chandra\ and the \xmm\
data, we also analyzed the  MOS and PN spectra of the 1-2.5' region
reduced with the latest
available calibration information
(see Section \ref{spectral} for details, Figures \ref{1mekal}-\ref{1mekal_0.5-4}
and Tables \ref{table-1mekal}-\ref{table-mekal-mekal} for results). Such a comparison indicates clearly that
both MOS and PN detect a behavior similar to the ACIS spectra: the spectra are not consistent
with a single-temperature model, and feature residuals at both low and high energy.
The addition of a power-law or of a low-temperature
thermal model results in significant improvements to the $\chi^2$, as was the case for 
the \chandra\ data.

Differences between the three instruments (ACIS, PN and MOS) 
remain, as the best-fit hot gas temperatures
from \xmm\ are somewhat lower than \chandra's, and the PN excess is $\sim$10\% higher than
those in ACIS and MOS. Since a wrong hot gas temperature may affect
the detection of the excess emission, we also perform a joint fit between all three instruments
(bottom right panel in Figures \ref{1mekal}-\ref{1mekal_0.5-4}). This exercise
indicates that even using \xmm-driven temperatures, the \chandra\ data
are not satisfactorily fit by a single-temperature emission model. 
Comparison of the best-fit non-thermal models (Table \ref{table-mekal-po})
between the three instruments indicates that a power-law of
index $\alpha \sim$ 1.8 is a viable model for both \chandra\ and \xmm\
data of \a3112. The two-temperature model (Table \ref{table-mekal-mekal})
also provides a significant reduction in $\chi^2$ for all datasets,
although the three instruments do not agree on the temperature of the warm phase.

One of the major sources of uncertainty in the ACIS calibration at low
energies is the presence of a contaminant on the optical blocking filter
of \chandra.~\footnote{See, for example, 
http:/cxc.harvard.edu/cal/Acis/Cal\_prods/qeDeg/index.html.} 
At present, the \chandra\ calibration team has developed a model
for the contaminant with an estimated
uncertainty on its optical depth of 10\% at 0.7 keV. 
Such an uncertainty will not be sufficient to explain the
20\% soft X-ray residuals shown in this paper.
The contaminant
contains elements with absorption edges in the 0.5-0.7 keV range, and
we estimated that 
an optical depth of the contaminant that is higher by a factor of 30\%
is necessary in order to explain the soft X-ray residuals present in this
\a3112\ observation. The fact that the \xmm\ data show residuals 
of similar nature and in comparable amount (and even in larger amount
according to the PN data) argues against such an instrumental nature of
the \chandra\ excess.

\section{Background subtraction and projection effects}
\label{background-projection}
\subsection{Background subtracion}
\label{background}
\begin{figure}[!h]
\begin{center}
\includegraphics[width=2.2in,angle=-90]{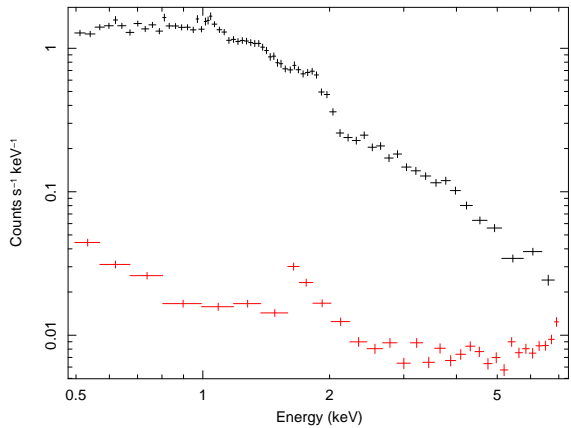}
\end{center}
\caption{Source spectrum of the 1-2.5' region (black) and blank-sky background (red) used for background
subtraction. 
\label{back}}
\end{figure}

We used the blank-sky datasets provided with the CIAO
software for the purpose
of background subtraction, as described in Section \ref{reduction}.
The quiescent X-ray background present in the data has
two main components: a stable particle background, and a
variable sky signal. The former is present in the blank-sky dataset, and therefore
accurately removed from our cluster data. The latter is a soft X-ray background which varies 
with position in the sky, and is the dominant source of background at energies $\leq$ 1 keV. 
The background in this \a3112\ observation is $<$4\% at soft energies (0.5-2 keV), 
and $\sim$25\% at energies 5-7 keV (Figure \ref{back}).

Our choice of investigating only the central -- and brightest --  regions of 
the cluster (excluding the core) was motivated precisely by the 
presence of this variable soft
X-ray background component, and the need to minimize its effects. 
If the  $\sim 20$\% soft X-ray excess 
was due to an anomalous background in this observation, it  
is required that the background exceed the blank-sky estimate
by a factor of 5. 
This exceeds the observed variability of the soft X-ray sky background 
by more than 1 order of magnitude,
which  is $\sim$25-50\% in the $\sim0.5-1$ keV band
\citep[e.g.,][]{bonamente2005}. 

In the high-energy band ($E \geq4$ keV), the diffuse
sky background is expected
to be negligible \citep{markevitch2003}, and the 
X-ray background is of detector origin. We established
 that the background subtraction
in this band was accurate by ensuring that the signal at high energy (E$\geq$10 keV), where
\chandra\ has no effective area to detect photons, 
was consistent between the \a3112\ observations
and the blank-sky observations, thereby resulting in a null background-subtracted
spectrum at those energies. The fact that the background is not
responsible for the apparent excess of hard photons
(Figures \ref{1mekal} and \ref{1mekal_0.5-4}) 
was also established by performing a fit to the spectrum in which the instrumental
background was artificially increased by 20\%; this test
resulted in no significant changes
to the fit parameters or the high-energy residuals.
\citet{markevitch2003} also shows that the high-energy background usually
does not vary between observations by more than a few percent.

\subsection{Projection effects \label{projection}}
Most clusters feature a radial temperature profile that decreases at large radii,
as found, e.g., by \citet{vikhlinin2006,vikhlinin2005}.
The temperature profile is typically flat over the range $\sim 0.1-0.3$ $r_{500}$,
where $r_{500}$ indicates the radius within wich the mean cluster density is
500 times higher than the critical density.
We examined whether the soft excess signal in the 1.0--2.5' annulus analyzed in this 
paper could be due to 
projection effects of gas at different temperatures along the line of sight.
For this purpose, we modeled the three-dimensional temperature profile of \a3112\ using 
the average model found in the analysis of several \chandra\ clusters
\citep[Eq. 9 in][]{vikhlinin2006}.
Using a mean temperature of 5 keV and a redshift of z=0.075, 
we find that r$_{500}\simeq 1.1$ Mpc, or $\sim$12' (assuming the concordance cosmology),
and therefore the inner radius of 1' corresponds to approximately $0.1 r_{500}$. 
For the gas density distribution we used the $\beta$ model with parameters obtained from \rosat\ PSPC data
\citep{vikhlinin1999}, i.e., $\beta$ = 0.63 and r$_{core}$ = 1.0'.
We divided the cluster in concentric shells of 0.5' width and assigned each shell with density 
and temperature values given by the above models. We then intersected the spherical shells with a hollow 
cylinder of inner and outer radii of 1.0 and 2.5', representing our line of sight to 
\a3112, and 
computed the volume of these portions of shell using exact analytical formulas. 
Finally, we calculated the emission measure of 
each shell at different temperatures along the line of sight. 

\begin{figure}[!h]
\includegraphics[width=2.5in,angle=-90]{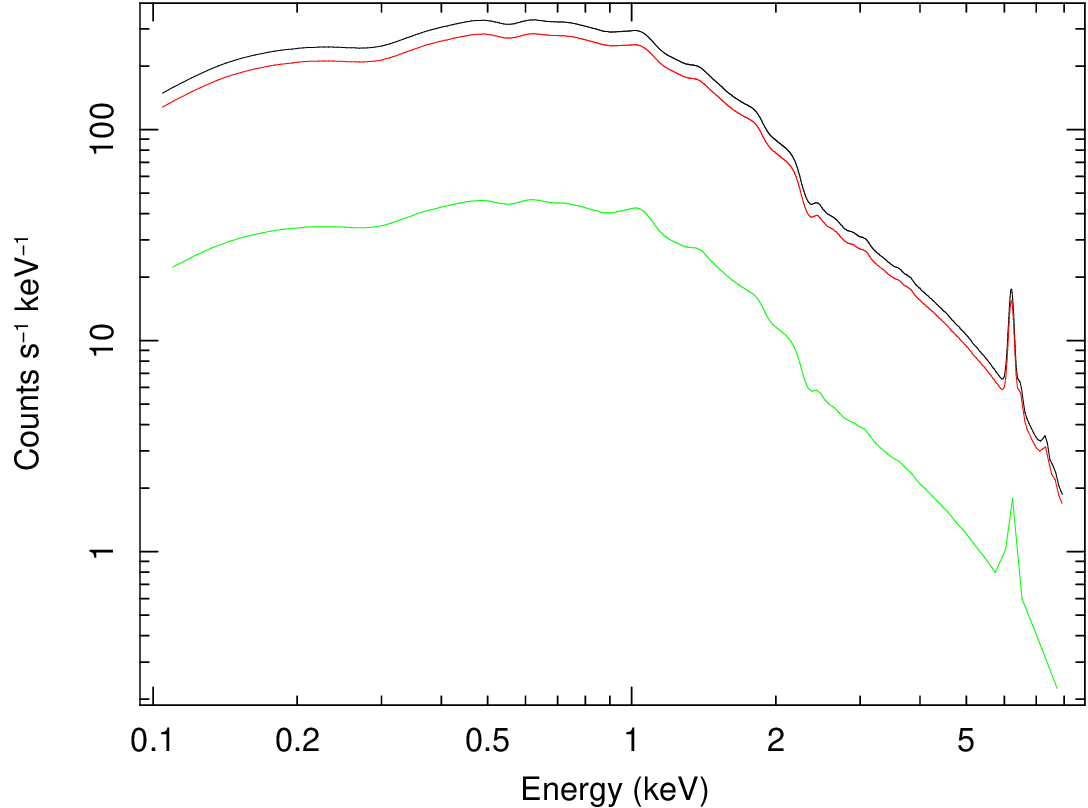}
\includegraphics[width=2.5in,angle=-90]{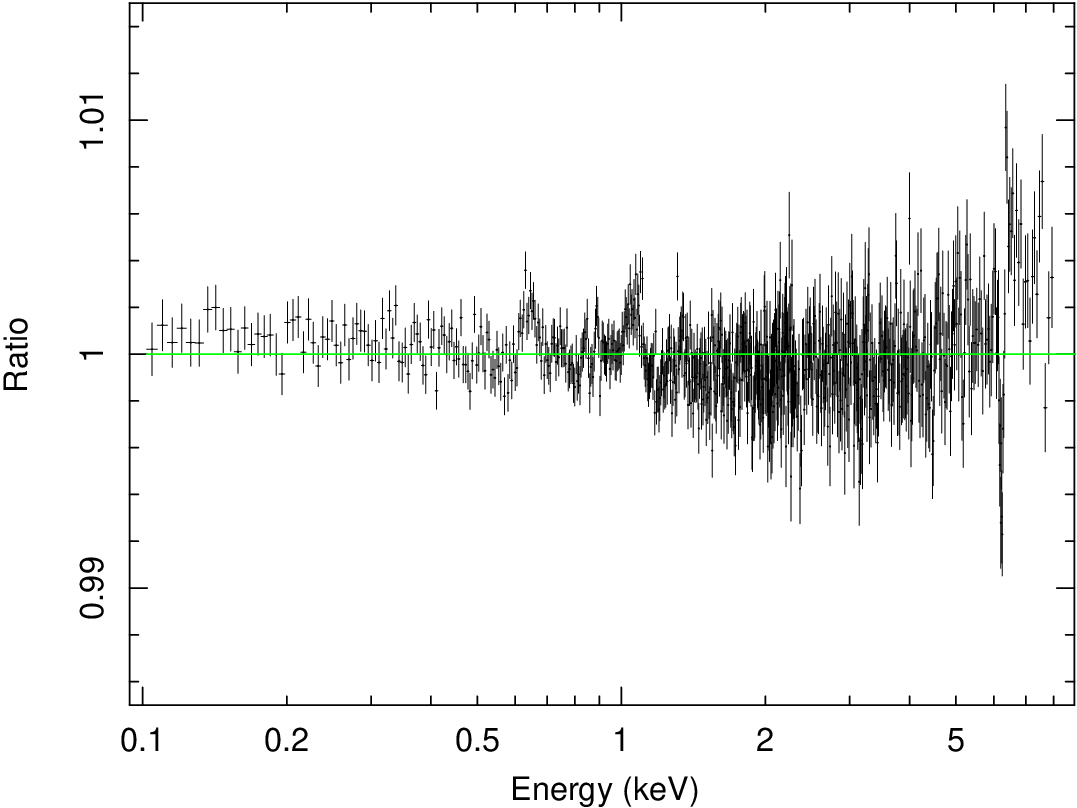}
\caption{Simulated \xmm-PN spectra from a 3D model of the cluster
atmosphere of \a3112. Left: In black: composite spectrum from all
spherical shells (0.1-1 $r_{500}$)
projected onto the 1-2.5' annulus; in red: spectrum from regions within
the isothermal portion of the cluster (0.1-0.3 $r_{500}$), projected onto
the 1-2.5' annulus; in green: spectrum from  outer regions only (0.3-1 $r_{500}$), again
projected onto the 1-2.5' annulus.
Right: residuals to an isothermal fit of the
composite spectrum (notice that each major tickmark is
1\% deviation). See Section \ref{projection} for further information.
\label{proj}}
\end{figure}

We found that $\sim$90\% of the emission in the projected 1.0--2.5' 
region of \a3112\ originates from three-dimensional distances of  
0.1--0.3 r$_{500}$ from the cluster center, where the gas temperature is nearly isothermal, i.e. varies by less 
than 10\%. Due to the dominance of the isothermal component the effect of the 
projected lower temperatures is expected to be small, as we further investigate
 in the following.
Using the emission measures and temperatures of our three-dimensional model described above,
 we simulated PN spectra (the most sensitive instrument available) for different portions
of the line-of-sight: 
(a) from a 3D radial range of 0.1--0.3 r$_{500}$ (inner isothermal region) , 
(b) from a radial range 0.3--1.0 r$_{500}$ (outer region) and (c) a radial
range of 0.1--1.0 r$_{500}$ (full region, assuming that the X-ray emission extends
to $r_{500}$);
all spectra were projected onto the 1-2.5' annulus. 
The best-fit temperature of the emission originating from the outer region is 20\% lower than 
that in the inner isothermal region, and its emission measure is 
only $\sim$10\% of that of the  hotter one (Figure \ref{proj}).
The full-region spectrum is fitted perfectly with a single temperature model with 
a best-fit temperature 3\% lower than that 
within the inner isothermal region (Fig. \ref{proj}).
There are no soft or hard X-ray residuals, even when considering 
the channels at the lowest energy of 0.1 keV, 
and thus the projection of different temperatures along the 
line of sight does not explain the soft excess detected in the 1.-2.5' region of \a3112. 

\section{Interpretation of the excess emission}
\label{interpretation}
Having  ruled out the \HI\ column density and the background as sources of the
excess emission, and addressed other sources
of systematic errors,
we now turn to the physical interpretation of the  excess emission,
as in the \xmm\ observation of \as1101 of \citet{bonamente2005}. 

\subsection{Non-thermal interpretation}
Relativistic electrons in the intergalactic medium
will cause cosmic microwave background (CMB) photons to Compton-scatter into the
X-ray band (the so-called inverse-Compton scattering). 
The Lorentz factor $\gamma$ of the electrons is related
to the observed energy $E$ (0.5-7 keV) of the photons via
$E = 75 \cdot \left( \gamma/300\right)^2$ eV,
\citep[e.g.,][]{sarazin1988}.
Assuming that each annulus is representative of a spherical shell,
one can calculate the pressure of the thermal and of the non-thermal
component. For the hot gas pressure, $p=n k T$, the 
number density $n$ is
estimated from the measured normalization $N$ of the spectrum,
$N \propto \int n^2 dV$, where $V$ is the volume of the 
spherical shell.
The non-thermal pressure is calculated as $p_{\rm nt}=\frac{1}{3}u$, where $u$ is
the energy density of the relativistic electrons,
 calculated as

\begin{equation}
u=\frac{1}{V} \cdot 8\times 10^{61} \cdot \left(\frac{L_{\rm nt}}{10^{42} \; \rm erg\; s^{-1}} \right) \left(\frac{3-\mu}{2-\mu}\right)
\left(\frac{\gamma_{\rm max}^{2-\mu}-\gamma_{\rm min}^{2-\mu}}{\gamma_{\rm max}^{3-\mu}-\gamma_{\rm min}^{3-\mu}}
\right) \;\rm erg\; cm^{-3}
\label{u}
\end{equation}
in which $L_{\rm nt}$ is the unabsorbed non-thermal luminosity in the 0.5-7 keV band,
$\mu$ is the electron differential number index ($dN_e/dE \propto E^{-\mu}$)
related to the observed spectral power-law index by $\mu=-1+2\alpha$.
The results are that the non-thermal pressure
accounts for a small fraction ($\sim 7$\%) of the hot gas pressure
(Table \ref{table-res}).

The relativistic particles, or cosmic rays, could be provided by
jets of a central active galaxy, then transported outwards while undergoing
in-situ second order Fermi acceleration by turbulent Alfv\'{e}n waves
\citep{lieu2007}.
If the acceleration of relativistic electrons occurs at diffuse shocks
\citep[e.g.,][]{bell1978a,bell1978b}, then a typical electron spectral index is $\mu \simeq 2.5$, 
corresponding to a photon index of $\alpha\simeq 1.75$, as observed 
by both \chandra\ and \xmm.
A steepening of the power-law index towards larger radii is naturally
explained as a result of radiative losses \citep{lieu1999b}.
Presence of non-thermal phenomena in the core of \a3112\ are also confirmed by the 
low-frequency radio emission associated with the central galaxy, and with a double-tailed
source within $\sim$30" of the cluster's center \citep{takizawa2003}.

\begin{deluxetable}{ccccc}
\tablecaption{Non-thermal and thermal interpretation of the excess emission\label{table-res}}
\startdata
\hline
  \multicolumn{2}{c}{NON-THERMAL} &  \multicolumn{3}{c}{THERMAL} \\
  \multicolumn{2}{c}{ } & \multicolumn{2}{c}{Intra-cluster gas model} & \multicolumn{1}{c}{Filaments model}\\
  \multicolumn{2}{c}{\hrulefill} & \multicolumn{2}{c}{\hrulefill} &  \multicolumn{1}{c}{\hrulefill}\\
$n_{\rm hot}$ & $p_{\rm nt}/p_{\rm thermal}$~\tablenotemark{(a)} &  $n_{\rm hot}$& $n_{\rm warm}$ & $n_{\rm fil}$ \\
    ($10^{-3}$ cm$^{-3}$) &  & ($10^{-3}$ cm$^{-3}$) & ($10^{-3}$ cm$^{-3}$) & ($10^{-3}$ cm$^{-3}$)\\
\hline
 2.4 & 7.2\% & 3.3 & 1.3 $f^{-1/2}$ & 0.70 $(L_{fil}/1\;\rm Mpc)^{-1/2}$\\
\enddata

\tablenotetext{a}{The non-thermal pressure $p_{nt}$ is calculated from the luminosity $L_{42}$ of
Table \ref{table-mekal-po}, using Equation \ref{u}.}
\end{deluxetable}
\subsection{Thermal interpretation \label{thermal}}
If the excess is of thermal nature, the mass of the warm phase
can be estimated for two simple geometries,  one (a) in which 
the warm gas co-exists with the hot gas in spherical shells, the other (b) in which the
gas is in diffuse filaments ({\it \'{a} l\'{a}} \citealt{cen1999})
of fixed density or length, projected towards the cluster.

(a) In the first case, the warm gas may be clumped with a volume filling factor
$f=V_{\rm eff}/V$, where $V_{\rm eff}<<V$ is the effective volume occupied by the gas,
and the detected emission integral is
$I = \int n^2 f  dV$. The gas density $n$ can therefore be estimated for a fixed
volume filling factor $f$, $n \propto f^{-1/2}$ (Table \ref{table-res}). 
The ratio of warm-to-hot gas mass is given by 
${M_{\rm warm}}/{M_{\rm hot}}= f^{1/2} \left({I_{\rm warm}}/{I_{\rm hot}}\right)^{1/2}$, 
corresponding to 39$\pm^{47}_{9}$ f$^{1/2}$ \%;
therefore, the warm gas may account for a significant 
fraction of the cluster's total baryon mass,
depending on its filling factor. 
Alternatively, the volume filling factor can 
be estimated directly by requiring that the warm and hot gas are in pressure equilibrium,
$p=n_{\rm hot} kT_{\rm hot}=n_{\rm warm} kT_{\rm warm}$. In this case
one can show that the ratio of warm-to-hot gas mass in each annular region
is ${M_{\rm warm}}/{M_{\rm hot}}=\left({I_{\rm warm}}/{I_{\rm hot}}\right)^{3/2} 
\left({T_{\rm warm}}/{T_{\rm hot}}\right)$. This results in a warm-to-hot gas mass 
ratio of 0.35$\pm^{0.72}_{0.17}$\%.
The putative
warm gas will have a cooling time that is shorter than the Hubble time,
thereby requiring a replenishment or heating mechanism in order to
be sustained. The cooling time can be estimated using the isobaric cooling formula
of \citet{sarazin1988}.
Assuming a diffuse warm gas of filling factor $f=1$, 
the cooling time is $1.9$ Gyrs;
if the warm gas is denser because of a smaller filling factor, or because it is 
in pressure equipartition with the hot gas, the cooling time is further reduced.

(b) In the second case, the volume occupied by the gas is $A\cdot L$, where
$A$ is the area of the annulus and $L$ the filament's length along
the line of sight, and the detected emission integral becomes
$I = \int n^2 (A dL)$.
In this case, one needs to fix either the density $n$ or the length $L$ in order to
interpret the detected emission integral $I$. In this paper, we assume
filaments of a fixed fiducial length of 1 Mpc, 
and estimate accordingly the density of the
warm filaments projected towards the cluster (Table \ref{table-res}). 
The density derived in this fashion scales 
as $n \propto L^{-1/2}$, i.e., if the filaments are 10 Mpc long instead, the density will be 
reduced by a factor of $\sim 3$. 
The density and length of the putative filaments according to this interpretation of the
soft excess are orders of magnitude larger that predicted by simulations
\citep[e.g.][]{cen1999,dave2001}, similar to the case of the soft excess in the \xmm\ observation 
\as1101\ \citep{bonamente2005}.

\subsection{Effects on the hot intra-cluster medium}

The results of Tables \ref{table-1mekal}, \ref{table-mekal-po} and
\ref{table-mekal-mekal}  show that the presence 
of an undiagnosed additional component in X-ray spectra, regardless of
its origin, has an impact on the
determination of the temperature and metal abundance of the hot gas.
The temperature is affected by the presence of the excess component,
which, if properly modeled with either a thermal or non-thermal
model, causes a systematic shift in the
measured $T$ by up to $\sim$25\%  in these observations of \a3112.
The measurement of chemical abundances will also experience a systematic
change towards larger values. In 
particular, if the
excess is of non-thermal origin, the
result is that of nearly Solar abundances \citep[as in the Perseus cluster,][]{sanders2006}, instead of
significantly sub-Solar as usually measured \citep{degrandi2001}.


\section{Discussion and Conclusions}
\label{conclusions}
The  \chandra\ ACIS-S observations of \a3112\ analyzed in this paper
feature an excess of X-ray 
photons which cannot be attributed to
uncertainties in the Galactic \HI\ absorbing column, or to the X-ray
background emission. 
The excess emission is equally well fit by  a non-thermal power-law model,
and by a thermal model of $\sim 0.2-0.7$ keV temperature and null
metal abundance. The excess is similar to that detected using
\xmm\ data \citep[][and this paper]{nevalainen2003}.

Both interpretations point to additional physical mechanisms at work
in galaxy clusters.
The thermal interpretation of the excess is inconsistent
with emission from diffuse filaments {\it \'{a} l\'{a}} \citet{cen1999}, and indicates
that the putative warm gas may be as massive as the hot gas. 
The non-thermal interpretation, on the other hand, suggests that
a significant fraction of the cluster's X-ray emission 
may be associated with non-thermal processes, according
to the original proposal of \citet{Felten1966}, and not with the
well-known hot gas at $T\sim 10^8$ K.
If a substantial fraction of the X-ray emission of some clusters
is of non-thermal origin, 
it may also explain the less-than-expected Sunyaev-Zel'dovich effect that emerged
from a comparison of WMAP and X-ray data for a large sample of nearby 
clusters \citep{lieu2006,afshordi2006}.
In fact, for equal pressure of thermal and
relativistic electrons, the Sunyaev-Zel'dovich decrement due to the latter
is much less than the former \citep[e.g.,][]{ensslin2000}.

Does our detection then constitute a {\it soft} or a {\it hard} excess? From
a pure data analysis viewpoint, the fact that
a two-temperature fit to the  spectra shifts the best-fit temperature
of the hot phase to higher temperature, and requires the introduction
of a softer component, cannot alone be considered
a proof that the data-model mismatch is due to a soft excess of thermal nature.
In fact, the non-thermal model has same goodness of fit as the 
two-temperature thermal model.
Moreover, the intrinsic paucity of hard X-ray photons compared to soft
X-ray photons is such that the latter drive the spectral fit, and therefore
a {\it bona fide} power law component may be confused for a low-temperature
thermal emission, at the resolution of these \a3112\ observations.
From an interpretational point of view, the presence of
non-thermal emission from radio observations, and the
difficulties with the thermal interpretation of the excess emission 
(Section \ref{thermal}), point at a non-thermal
origin of the emission. We therefore argue in favor of a non-thermal origin 
of the excess, and that the phenomenon is both a soft and hard excess or, simply,
an excess emission throughout the X-ray bandpass.

The presence of this X-ray excess emission also affects the determination
of the hot gas parameters. Understanding the origin of this  excess emission
therefore promises not only the discovery of new physical phenomena in galaxy clusters, 
but also a better knowledge of the hot intergalactic medium. 
\vspace{2cm} 

\acknowledgements

The authors thank M. Markevitch  and A. Vikhlinin for valuable suggestions
on the \chandra\ data analysis, and the anonymous referee for several 
suggestions that led to significant improvements to the manuscript. 
JN acknowledges the support from the Academy of Finland.

\bibliographystyle{apj}

\end{document}